\documentclass[acmlarge,screen]{acmart}

\usepackage{hyperref,booktabs,tabularx,multirow}
\usepackage[flushleft]{threeparttable}


\setcopyright{acmcopyright}
\acmJournal{IMWUT}
\acmYear{2018} 
\acmVolume{2} 
\acmNumber{1} 
\acmArticle{8}
\acmMonth{3} 
\acmPrice{15.00}
\acmDOI{10.1145/3191740}

\begin{document}

\title{Mobile Device Type Substitution}

\author{Benjamin Finley}
\affiliation{%
  \institution{Aalto University}
  \department{Department of Communications and Networking}
  \streetaddress{Konemiehentie 2}
  \city{Espoo}
  \postcode{02150}
  \country{Finland}}
\email{benjamin.finley@aalto.fi}
 
\author{Tapio Soikkeli}
\affiliation{%
  \institution{Verto Analytics}
  \streetaddress{Innopoli 1, Tekniikantie 12}
  \city{Espoo}
  \postcode{02150}
  \country{Finland}}
\email{tapio.soikkeli@vertoanalytics.com}

\begin{abstract}
Mobile users today interact with a variety of mobile device types including smartphones, tablets, smartwatches, and others. However research on mobile device type substitution has been limited in several respects including a lack of detailed and robust analyses. Therefore, in this work we study mobile device type substitution through analysis of multidevice usage data from a large US-based user panel. Specifically, we use regression analysis over paired user groups to test five device type substitution hypotheses. We find that both tablets and PCs are partial substitutes for smartphones with tablet and PC ownership decreasing smartphone usage by about 12.5 and 13 hours/month respectively. Additionally, we find that tablets and PCs also prompt about 20 and 57 hours/month respectively of additional (non-substituted) usage. We also illustrate significant inter-user diversity in substituted and additional usage. Overall, our results can help in understanding the relative positioning of different mobile device types and in parameterizing higher level mobile ecosystem models.
\end{abstract}


\maketitle

\renewcommand{\shortauthors}{B. Finley and T. Soikkeli}

\section{Introduction}
The smartphone is the most iconic and ubiquitous mobile device of the past decade. However, the smartphone is far from the only mobile device available to consumers given the introduction of device types such as the tablet, smartwatch, and the introduction of mobile connectivity to the personal computer (PC) (primarily the laptop). In fact, many device manufacturers now have products in all of these mobile device lines.

Given this diversity, understanding how certain device types may substitute for other device types (specifically a shift of usage to a new device) or even prompt additional (non-substituted) usage is important for, as an example, understanding the relative positioning of these device types in the mobile ecosystem. Yet despite this importance, few studies have explicitly analyzed device type substitution \citep{matthews2009,shmorgun2013,muller2015,finley2016,finley2017a}. Furthermore, existing studies have either relied on survey or user completed diary methods\footnote{These methods are susceptible to recall bias \citep{dereuver2012}.} \citep{matthews2009,shmorgun2013,muller2015} or neglected matching methods to control for covariate imbalances between compared groups \citep{finley2017a}. 

In this work, we examine device type substitution through an analysis of multidevice (smartphone, tablet and PC) and multiplatform (Android, iOS, etc.) usage data from a large US based user panel. Importantly, the data collection method is device-based (users installed custom device-based monitoring applications), and we use a robust matching method. Specifically, we use a coarsened exact matching (CEM) strategy to create matched groups where the major difference between the groups (known as the treatment) is ownership of a particular device type. Then regression allows for the estimation of the effect of that device type ownership on usage of another device type.

Through our analysis, we test several different device type substitution hypotheses partially informed by prior research \cite[Section 5.2]{finley2016} \cite[Section 5.4]{finley2017a}. The tested hypotheses are as follows:

\begin{itemize}
  \item[] H1: Tablet ownership decreases smartphone usage
  \item[] H2: Tablet ownership decreases PC usage
  \item[] H3: PC ownership has no effect on smartphone usage
  \item[] H4: Tablet ownership increases total device usage\footnote{Total device usage is the sum of usage of all smartphone, tablet, and PC devices of the user.}
  \item[] H5: PC ownership increases total device usage
\end{itemize}

These hypotheses, if supported, can help provide support for the following more general statements:

\begin{itemize}
  \item[] S1: Tablets are a partial substitute for smartphones
  \item[] S2: Tablets are a partial substitute for PCs
  \item[] S3: PCs are not a partial substitute for smartphones
  \item[] S4: Tablets prompt additional (non-substituted) usage
  \item[] S5: PCs prompt additional (non-substituted) usage
\end{itemize}

The remainder of this work is organized as follows. Sections \ref{background} details background information including a summary of related work and an overview of the dataset. Section \ref{preprocessing} describes the preprocessing steps including several data issues and the robust matching method. Finally, Section \ref{results} reports the analysis results including hypothesis testing, Section \ref{discussion} discusses theoretical reasoning, limitations, and examples of implications, and Section \ref{conclusions} concludes the work.

\section{Background} \label{background}
This section describes the background information including related work and a dataset overview.
\subsection{Related Work} \label{sect_rel_work}
Related work in mobile device type substitution is somewhat sparse. Studies can generally be classified into two broad categories based on the data collection methodology.

Several studies have used interview, survey, or electronic diary methods that ask users to recall their usage and device substitution behavior. 

\citet{matthews2009} studied device type substitution through semi-structured interviews of users. The results detailed that users substitute smartphones for laptops and vice versa in some situations (mainly personal usage rather than work usage) but that a majority of usage is instead additional. The interviews elicited a few substitution examples such as users deferring reading of full news articles until they could access their larger display laptops. \citet{shmorgun2013} analyzed device type substitution through a survey of 101 mobile device users dispersed through several different countries. The analysis detailed that users prefer specific device types for specific tasks even though the general usage volume of service types across device types is roughly equal. Finally, \citet{muller2015} used both surveys and a self-reporting electronic diary method to compare smartphone and tablet usage of 176 US-based mobile users. The comparison detailed that total device usage was highest among users with both smartphones and tablets, thus suggesting these devices prompt at least some additional usage (rather than simply substituted usage). 

Other studies have used data from device-based monitoring of mobile usage. 

\citet{finley2017a} studied device type substitution by comparing usage between a group of users with only smartphones and a group with both smartphones and tablets. The results indicated that tablet usage is partially substituted usage from smartphones and partially additional usage. The fraction of each of these cases was about half. Also, \citet{finley2016} explored device type substitution via correlation analysis of usage statistics of several different user groups (smartphone and tablet owners, smartphone and PC owners, etc.). In other words, instead of looking for the presence of substitution across groups, the work looked at the level of substitution within a group. The results suggested a significant substitution effect for users with both tablets and PCs; specifically the more applications and usage time on a tablet the less on a PC.

Finally, several recent mobile and internet related studies have similarly applied coarsened exact matching for reducing covariate imbalance before group comparison. 

For example, \citet{sandeep2017} used coarsened exact matching when analyzing the effect of developers providing free versions of paid mobile applications on paid applications adoption rate. Similarly, \citet{wen2017} applied coarsened exact matching when analyzing the effect of the platform owner (in this case Google) entering a specific area of the mobile application market (with native features or their own app). While \citet{cotten2014} applied coarsened exact matching in analyzing the effect of internet usage on depression in retired adults in the USA.

\subsection{Dataset} \label{datasets}
The primary dataset consists of one month of device usage data from a subset of active users of a large ongoing United States based user panel organized by Verto Analytics\footnote{\url{http://vertoanalytics.com/}}. Hereafter, we refer to the large on-going Verto panel as the general panel and the one-month subset of active users as the active dataset. 

Regarding the general panel, users are recruited online and given an initial recruitment survey that asks about the devices they own. Users are instructed to install custom monitoring applications to all of their applicable devices (in other words, the devices they report owning including smartphone, tablet, and PC). Monitoring applications are available for Google Android, Apple iOS, and Microsoft Windows. We discuss Apple macOS later. The monitoring applications log events such as, for Android, an application moving to the foreground of the device display. Only users that install monitoring applications to all their applicable devices are considered for the panel. All users are paid for participation. Also, users rarely install the monitoring applications to their work devices\footnote{Work devices generally do not permit users to grant the broad permissions the monitoring application requires.}.

For all devices, device usage time is calculated as the sum of device application sessions. For Android devices, application sessions are based on an application entering and then leaving the foreground of the display. Similarly for iOS devices, application sessions are based on a combination of display state and system calls that indicate an application entering the foreground of the display\footnote{The identification of the specific application on the foreground of the iOS device occurs through network traffic analysis, though identifying the specific application is not relevant for our analysis.} and then being replaced by a new application entering the foreground. Finally, for Windows devices, application sessions are based on an application window gaining and then losing focus.

For some applications the display is typically off during usage (for example, calling and music applications). For technical reasons we do not include this display-off usage. We discuss the implications of this further in Section \ref{limitations}. Concerning overlapping usage (between devices), we include such usage because our hypotheses are formulated in terms of total usage and overlapping usage is both possible and relatively common \citep{finley2017a}. We apply maximum session length timeouts of 3 hours for iOS/Android sessions and 8 hours for Windows sessions to limit sessions where the display is forced on continuously but the user may not be using the device. Finally, we exclude users that report owning an Apple PC since a monitoring application for Apple macOS is unavailable\footnote{For completeness, we note that Verto Analytics does currently have an Apple macOS monitor but did not at the time of the data collection.} and we do not have PC usage information for these users.

The active dataset consists of one month (November 2016) of device usage data from 5158 active users. The dataset additionally includes demographic and smartphone technographic data for these users. The dataset does not include personally identifying information. 

\begin{figure}[!htb]
\centering
\includegraphics[width=4.7in]{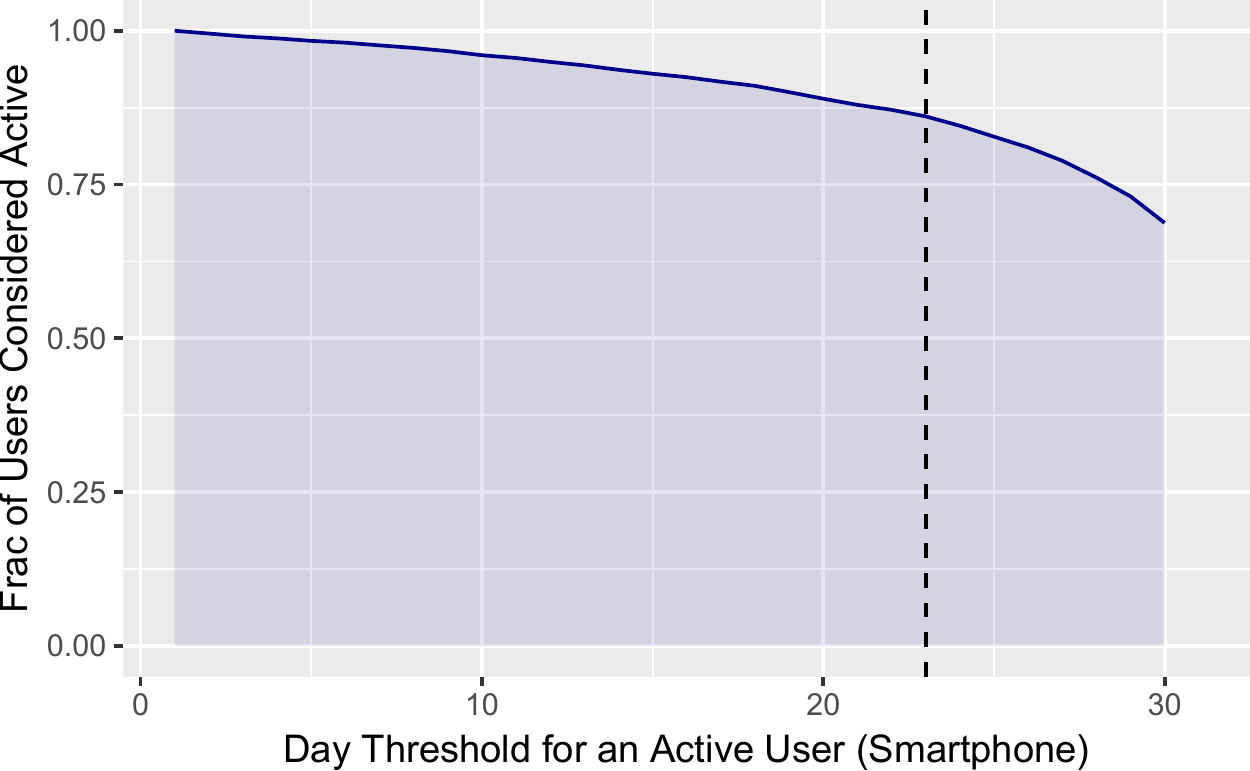}
\caption{Fraction of users considered active given different day thresholds for smartphone. The dashed line denotes the selected threshold of 23 days.}
\label{day_thresh}
\end{figure}

We define a user as active if they used their smartphone on at least 23 days of the month. This day threshold represents a trade-off between the sample size (number of analyzed users) and sample definition (ensuring analyzed users are truly active panel members throughout the month\footnote{In other words, ensuring that users that essentially drop out of the panel are not included.}). Figure \ref{day_thresh} illustrates the normalized number of users considered active with different day thresholds including the selected threshold of 23 days. We do not define active thresholds for tablet and PC usage because the notion of activeness is less clear for these device types that are not necessarily everyday drivers. We also note that even our superset of both active and non-active users (about 8000 users total) is a subset of the full general panel.

In terms of representativeness, general panel recruitment is performed with the purpose of obtaining a nationally representative panel and thus the general panel is relatively diverse. However, all opt-in panels inherently use non-probability sampling and thus representativeness is a concern. We refer to \citet{hays2015} for a thorough discussion of non-probability internet sampling. Furthermore, as mentioned, the recruitment process uses a recruitment survey that screens potential users to improve the demographic and technographic match between the accepted users and the population (known as a quota-sampling approach).

For reference, we provide a summary of demographic data for active dataset users along with demographic data for US smartphone users in general in Table \ref{panel_demos}. There are several demographic discrepancies. We do not attempt to calculate and use, for example, raking weights to remove these discrepancies partly because our matching package does not support such weights. In any case, we agree with \citet{church2015} that these types of studies are mainly about the panel populations themselves and the value is primarily in allowing researchers and the community to compare and contrast different experiences with different user populations to build a comprehensive understanding of the variety of mobile users and their unique behaviors.

\begin{table}[!htb]
\centering
\begin{threeparttable}
\caption{Comparison of Demographics between Active Dataset Users and US Smartphone Users}
\label{panel_demos}
\begin{tabular}{lrr}
\toprule
{\bf Demographic} & {\bf Active Users} & {\bf US Smartphone Users\tnote{a}}\\ \midrule
Mean Age (Years)\tnote{b} & 35.24 (12.29) & 42.19 (15.72)\\
Gender (\% Male) & 30.57  & 50.15\\ 
Marital Status (\% Married)\tnote{c} & 42.94 & 56.98 \\
Employment Status (\% Employed)\tnote{d} & 59.09 & 72.15 \\
HH\tnote{e}\, Income (\% \textless \$50K) & 68.15 & 44.58 \\ 
Mean HH\tnote{e}\, Size & 3.00 (1.58) & 3.12 (1.58) \\ 
Mean Children in HH\tnote{e} & 0.88 (1.19) & 0.75 (1.10)\\
Race (\% White) & 74.16 & 74.00 \\
Ethnicity (\% Hispanic) & 10.19 & 15.24 \\
\bottomrule
\end{tabular}
\begin{tablenotes}
    \scriptsize
    \item[a] US smartphone user demographic data is from Pew Research survey (n=3015, subpop with smartphone n=2310) \citep{pew2016} from September-November 2016. The survey utilizes weighting to population parameters of census data to create nationally representative results (refer to \citet{pew2017}). Verto Analytics also performs its own national surveys, we utilize the Pew Research survey only for brevity.
    \item[b] All mean values also include standard deviations
    \item[c] Married includes responses for both married and domestic partnership so that the definition matches the coarsening used in CEM (see Table \ref{summary_coarsening_categ})
    \item[d] Employed includes responses for full-time, part-time, and self employment so that the definition matches the coarsening used in CEM (see Table \ref{summary_coarsening_categ})
    \item[e] Household
\end{tablenotes}
\end{threeparttable}
\end{table}

For completeness, we note that the demographic data includes age, gender, marital status, employment status, household income, household size, children in household, race, and ethnicity. While the smartphone technographic data includes smartphone model, platform, display size, display pixel density, display-to-body ratio, and RAM. The technographic data for each smartphone model was collected from public sources such as \url{gsmarena.com}. The technographic data is included (in addition to the demographic data) in the matching method and analysis under the assumption that smartphone technographics are a rough proxy for the technical sophistication of the user. Therefore, the technographic data complements the demographic data since such technical sophistication is unlikely to be predictable by demographics alone (see the correlations between demographics and technographics in Figure \ref{covariate_correlations}). We detail a summary of technographic data and device type ownership for active dataset users in Table \ref{panel_technos}.

\begin{table}[!htb]
\centering
\begin{threeparttable}
\caption{Technographics and Device Type Ownership of Active Dataset Users}
\label{panel_technos}
\begin{tabular}{lr}
\toprule
{\bf Technographic} & {\bf Active Users} \\ \midrule
Smartphone Display Size (Inches)\tnote{a} & 4.98 (0.51) \\
Smartphone Display Density (PPI)\tnote{b} & 380.74 (126.22) \\ 
Smartphone Display-to-Body Ratio (\%) & 68.76 (5.08)\\
Smartphone RAM (GB) & 1.98 (0.95)\\ \midrule
{\bf Ownership} &  {\bf Active Users} \\ \midrule
Tablet Ownership (\% w/ Tablet) & 16.38\\ 
PC Ownership (\% w/ PC) & 50.72\\ 
Tablet and PC Ownership (\% w/ Tablet \& PC) & 7.33\\ 
\bottomrule
\end{tabular}
\begin{tablenotes}
    \scriptsize
    \item[a] All mean values also include standard deviations
    \item[b] Pixels per inch
\end{tablenotes}
\end{threeparttable}
\end{table}

\subsection{Hypotheses Formulation} \label{sect_hypothesis_formulation}
We briefly discuss the formulation of the hypotheses. All our hypotheses are formulated given the assumption that users have and use a smartphone (which is valid for our active dataset given that the definition of active is based on smartphone ownership and usage). Alternatively we could have formulated our hypotheses and/or selected our active dataset in several other ways\footnote{For example, we could have selected a different active dataset for each hypothesis. In the case of H2, for instance, the active dataset might consist of all users active with a PC or all users active with both a PC and tablet. However, as previously discussed, given that tablets and PCs are not necessarily everyday devices the notion of an active user is less straightforward.}. However, we believe that our smartphone centric focus is acceptable given the importance and ubiquity of the smartphone in the mobile ecosystem. Additionally, we do not believe that alternative formulations would significantly change our results.

Additionally, our hypotheses are formulated to analyze the effect of {\it ownership} (of one device type) on the {\it usage} (of another device type). These analyses are helpful in certain situations, for example, when the effect of acquiring a certain device type should be estimated. Alternative analyses, such as the effect of {\it usage} (of one device type) on {\it usage} of (another device type), as in Section \ref{correlation_approach}), are helpful in other situations. For example, when the effect of promoting usage of a certain device type should be estimated. In general we believe these analyses are complimentary. We look to perform more robust {\it usage} on {\it usage} analyses, beyond Section \ref{correlation_approach}, in future work.

\section{Preprocessing}\label{preprocessing}
This section describes the preprocessing steps including several data issues and the robust matching method.
\subsection{Missing Data}
In terms of missing data, 12 users of the active dataset (0.19\%) are missing smartphone technographic data. These missing values are related to users with very uncommon smartphones that have ambiguous model names. Given the very small amount of missing data, for simplicity we illustrate all analysis on complete case data. In other words, we exclude these users.

\subsection{Users with Multiple Devices of the Same Type}
A small fraction of users have multiple devices of the same device type\footnote{Specifically, for the active dataset, 6.3\% of users have multiple smartphones, 4.7\% of tablet users have multiple tablets, and 7.6\% of PC users have multiple PCs}, for example both a laptop and desktop computer. In calculating device type usage, we use a simple approach of summing all of a user's usage for devices of the same type. We also test an alternative approach of selecting one of the multiple devices at random, however this approach does not significantly change the analysis results. Therefore we only include the results of the summation approach.

\subsection{PC Subtypes (Desktops and Laptops)}
In the analysis we include both desktops and laptops in the PC device type under the assumption that the desktop and laptop sub-types are similar in terms of device type substitution. However, for robustness, we also test these subtypes separately in Section \ref{desktop_laptop_robustness}.

\subsection{Devices Shared within a Household}\label{shared_devices}
Devices, especially tablets and PCs, may be shared among several members of a single household \citep{muller2012}. For example, \citet{globalwebindex2017} reports that about 60\% of users share their tablet with at least one other user, though the extent of such sharing is not directly quantified. In terms of implications for our analysis, shared usage of tablets and PCs does not directly affect testing of H1 and H3 since these hypotheses use binary ownership/non-ownership variables for tablet and PC that would not be affected by shared usage. However, shared usage does affect the testing of H2, H4, and H5. Therefore, for robustness, we also test these hypotheses with only one person household users (thus lessening the possibility of device sharing) in Section \ref{device_sharing_robustness}.

\subsection{Matching Method of Coarsened Exact Matching}
In order to determine the effect of a treatment such as owning a tablet on the usage of another device type such as smartphone, the treatment group and control group should be as similar as possible in all other regards (i.e. concerning covariates\footnote{In our case these covariates are the demographic, techographic, and independent device usage variables.}). In this way, the treatment effect is isolated. There are several ex-post methods for creating these groups based on analysis of covariates. The most prominent of these methods is probably propensity score matching (PSM). However, PSM has several undesirable properties and alternative methods are often preferable \citep{king2015}.

In this work, we use coarsened exact matching (CEM) via the \texttt{cem} package (1.1.14) \citep{iacus2009} in R. A few of the advantages of CEM include giving the ability to control the amount of imbalance in the final matching through ex-post decisions (specifically the selected coarsenings), meeting the congruence principle\footnote{Specifically, the congruence principle specifies congruence between the data space and the analysis space. Refer to \citet[Section 4.2]{iacus2012} for further discussion.}, and being efficient in computational terms \citep{iacus2012}.

Table \ref{summary_coarsening_categ} details the coarsening of categorical covariates (aggregating of categorical levels) and Table \ref{summary_coarsening_contin} details the coarsening of continuous covariates (discretizing of continuous covariates). Whereas Table \ref{cem_results} details the results of the matching including the size of the resulting groups. The specific demographic coarsenings were primarily selected based on common coarsenings used in sociological work. Whereas, the specific technographic coarsenings were based on examination of the technographic covariate histograms and prior knowledge of common smartphone characteristics. 

For some covariates, such as race and ethnicity, an intuitive sociological coarsening is not apparent. In these cases, we take the approach of grouping together the covariate's smaller categories under the assumption that those users are somewhat similar. Alternatively we could use the covariate as is (without coarsening), which would discard many of those users from the matching, or we could remove the covariate from the matching method altogether. All these approaches make implicit assumptions, but in any case, given the small size of the categories the effect on our results should be small.

In terms of the actual matching procedure we use k2k matching with random selection. In k2k matching, within each stratum (combination of covariates) we randomly match a treatment and control sample (without replacement) until we exhaust either all treatment or all control samples\footnote{In other words, the maximum number of matchings in a stratum is the minimum of the treatment and control samples sizes}. The remaining non-matched treatment or control samples within the stratum are discarded.

An alternative approach is to randomly match treatment and control samples but to allow repeated matching of the same sample if the treatment and control sample sizes are different\footnote{In other words, the maximum number of matchings in a stratum is the maximum of the treatment and control samples sizes}. In this case, the subsequent matching will have weights for use in analysis. We also test with this alternative approach, however, the approach does not significantly change any of the analysis results. Therefore we only include the analysis results of the more straightforward k2k approach.

Finally, for illustration, Tables \ref{prematching_comparison} and \ref{postmatching_comparison} detail difference measures for covariates between treatment and control groups for M1 before and after matching, respectively. Notice that the differences between the treatment and control groups are diminished, though not completely removed. Therefore, the covariates are included in the final regression models to help control for the remaining differences.

\begin{table}[!htb]
\centering
\begin{threeparttable}
\caption{Summary of Coarsening of Categorical Covariates}
\label{summary_coarsening_categ}
\begin{tabularx}{15.5cm}{l>{\raggedright\arraybackslash}Xl}
\toprule
{\bf Covariate } & {\bf  Coarsened Categories} & {\bf Final Category }\\
\midrule
\multirow{2}{*}{Marital Status}   & Married, Domestic Partnership   & Married      \\ \cline{2-3}
                                  & Single, Divorced, Separated, Widowed  & Not Married  \\ 
\addlinespace[0.3cm]
\multirow{2}{*}{Employment Status}       & Employed - full-time, Employed - part-time, Self-Employed & Employed \\ \cline{2-3}
                                  & Homemaker, Unemployed and not looking for a job/Permanently Disabled/Housewife, Student, Currently Unemployed, Unemployed but looking for a job, Retired, Military\tnote{a}, Don't Know / Not Sure & Not Employed \\ 
\addlinespace[0.3cm]                                  
\multirow{3}{*}{HH Income (USD)} & \$15K \textless, [\$15K,\$20K), [\$20K,\$25K)  & Low-Income  \\ \cline{2-3}
                                  & [\$25K,\$30K), [\$30K,\$40K), [\$40K,\$50K), [\$50K,\$75K) & Middle-Income \\ \cline{2-3}
                                  & [\$75K,\$100K), [\$100K,\$150K), \textgreater \$150K & High-Income \\ 
\addlinespace[0.3cm]                                  
\multirow{2}{*}{Race}             & White & White \\ \cline{2-3}
                                  & Other, Asian, Black  & Non-White    \\ 
\addlinespace[0.3cm]                                  
\multirow{2}{*}{Ethnicity}        & Hispanic  & Hispanic   \\ \cline{2-3}
                                  & Not Hispanic, Don't Know / Not Sure   & Non-Hispanic \\
\bottomrule
\end{tabularx}
\begin{tablenotes}
    \scriptsize
    \item[a] Military is counted as not employed simply because military members are typically excluded from the definition of civilian employment. In any case, the number of military users is very small.
\end{tablenotes}
\end{threeparttable}
\end{table}

\begin{table}[!htb]
\centering
\begin{threeparttable}
\caption{Summary of Coarsening of Continuous Covariates}
\label{summary_coarsening_contin}
\begin{tabular}{ll}
\toprule
{\bf Covariate } & {\bf Intervals for Discretization} \\
\midrule
Age (Years)     & [18,25], [26,45], [46,82] \\ 
HH Size         & [1], [2], [3,13] \\ 
Children in HH  & [0], [1,9] \\ 
Display Size (Inches)   & [2.8,3.5], (3.5,4.5], (4.5,5.5], (5.5,6.5] \\ 
Diplay Density (PPI)\tnote{b}  & [132,200], (200,300], (300,400], (400,500], (500,800]\\
Display-to-Body Ratio (\%) & [0.34,0.5], (0.5,0.6], (0.6,0.7], (0.7,0.82] \\ 
RAM (GB)             & [0.15,0.6], (0.6,1], (1,2], (2,3], (3,4], (4,6] \\ 
Usage Times\tnote{a} & 6 equal width bins over range \\
\bottomrule
\end{tabular}
\begin{tablenotes}
    \scriptsize
    \item[a] Specifically smartphone, PC, and tablet usage times.
    \item[b] Pixels per inch
\end{tablenotes}
\end{threeparttable}
\end{table}

\begin{table}[!htb]
\centering
\begin{threeparttable}
\caption{CEM Matchings}
\label{cem_results}
\begin{tabular}{llll}
\toprule
{\bf Matching } & {\bf User Subset} & {\bf Treatment} & {\bf Users} \\ \midrule
M1 & All Users & Has Tablet & 514 \\
M2 & Has PC & Has Tablet & 178 \\
M3 & All Users & Has PC & 1218 \\
M4\tnote{a} & All Users & Has Tablet & 548 \\
M5\tnote{b} & All Users & Has PC & 1234\\
\bottomrule
\end{tabular}
\begin{tablenotes}
    \scriptsize
    \item[a] The difference between M1 and M4 is that M1 includes the PC usage time covariate in the matching while M4 does not include PC usage time because the dependant variable in H4 (total device usage) by definition includes PC usage time.
    \item[b] The difference between M3 and M5 is that M3 includes the tablet usage time covariate in the matching while M5 does not include tablet usage time because the dependant variable in H4 (total device usage) by definition includes tablet usage time.
\end{tablenotes}
\end{threeparttable}
\end{table}

\section{Results} \label{results}
In this section, we present and discuss the main results of our analysis.

\subsection{Basic Descriptive Statistics}
\begin{figure}[!htb]
\centering
\includegraphics[width=4.7in]{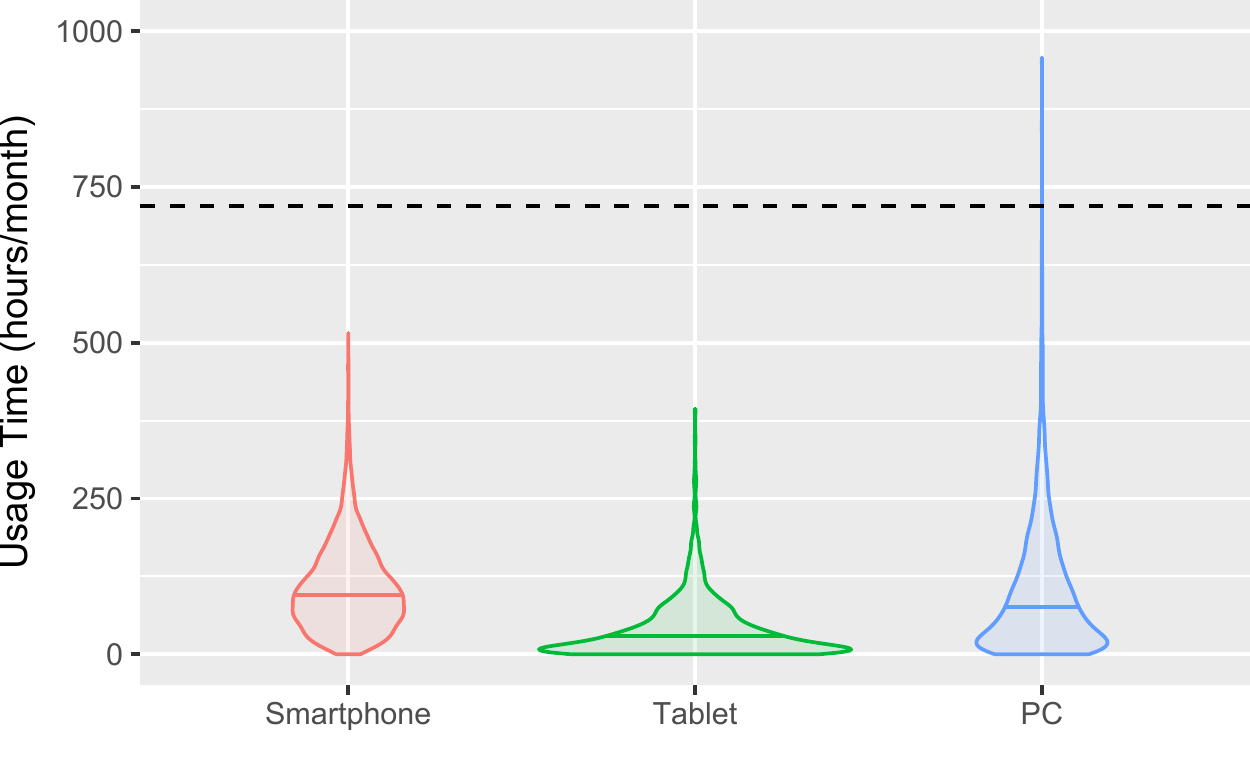}
\caption{Violin plot of usage time distributions by device type. Solid lines denote distribution medians and the dashed line denotes the total number of hours in the month.}
\label{device_type_comparison_violin}
\end{figure}

We first illustrate several basic statistics of the active dataset to help provide further context since many of the variables are non-normal. Figure \ref{device_type_comparison_violin} illustrates usage time distributions by device type as a violin plot. Interestingly, all three device types have quite different distribution shapes. Smartphone has the highest median usage time. However, PC has greater variation with a thicker and longer tail with significant outliers. 

\begin{figure}[!htb]
\centering
\includegraphics[width=4.7in]{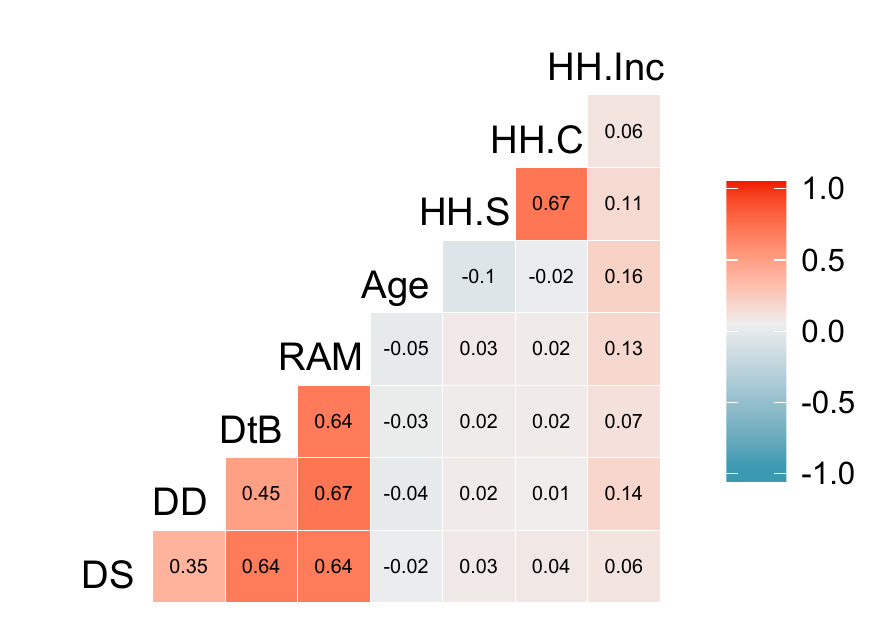}
\caption{Correlations (Kendall) between all numerical and ordinal covariates. (DS - Display Size, DD - Display Density, DtB - Display-to-Body Ratio, HH.S - Household Size, HH.C - Children in Household, HH.Inc - Household Income) Kendall correlation is used because household income is ordinal and Pearson correlation only supports numerical values. }
\label{covariate_correlations}
\end{figure}

Additionally, Figure \ref{covariate_correlations} details the Kendall correlations between all numerical and ordinal covariates. Interestingly, there are relatively few strong correlations between demographic and technographic covariates; though the moderately sized correlations that exist appear reasonable. For example, the positive correlation between HH income and display density suggests that high-income users generally have higher quality and more expensive devices (assuming that display density is a rough proxy for device quality and price\footnote{For completeness, we calculate the correlation between display density and price (as reported by gsmarena.com) for the 144 unique device models of users from \citet{finley2017b}. We find a strong positive Pearson correlation of 0.65.}).

\subsubsection{Hypotheses Testing}
To test the hypotheses, for each hypothesis we perform linear regression analysis over a matching and against an dependent variable as specified in Table \ref{testing_results}. However, before regression analysis, we perform several diagnostics to ensure validity.

Regarding multicollinearity, highly correlated covariates in a regression model can cause problems including inflated standard errors. Thus we perform multicollinearity diagnostics on the covariates using the \texttt{vif} command of the \texttt{car} package (2.1-4) \citep{fox2016} in R which reports generalized variance inflation factors (GVIF) for each covariate. First, we transform each GVIF to a factor comparable to non-generalized VIFs (denoted as a cVIF\footnote{The GVIFs are transformed to cVIFs via the equation $c=g^{\tfrac{1}{d}}$ where $g$ is the GVIF and $d$ is the degrees of freedom of the covariate \citep{fox2002}. The degrees of freedom for a numeric covariate is one and for a categorical covariate is the number of categories.}). We then use a step-wise elimination technique in which we remove the covariate with the highest cVIF over ten\footnote{A VIF cutoff of ten is a widely used guideline originally proposed in \citet{marquardt1970}.} until no remaining covariates have cVIFs over ten. Through this technique, we remove the covariate household size in H2 and no covariates in H1, H3, H4 and H5.

In terms of linear regression assumptions, the data in each matching generally does not meet the assumptions of normality of errors or homoscedasticity. These assumption violations are due primarily to the presence of significant outliers. Therefore, we use robust linear regression that is resistant to outliers and the mentioned assumption violations in general. Specifically, we utilize the robust linear regression command \texttt{lmrob} of the \texttt{robustbase} package (0.92-7) \citep{rousseeuw2009} in R. The command builds a linear regression model through an MM-type regression estimator \citep{yohai1987,koller2011}. This type of estimator has a high breakdown point of 50\% and high asymptotic efficiency. Specifically, in our case, the full estimator chain is an S-estimate, M-estimate, D-estimate (Design Adaptive Scale \citep{koller2011}), and another M-estimate as recommended in \citet{koller2017}. 

The estimator achieves robustness partly by down-weighting severe outliers (the resulting weights are known as robustness weights). In our case, this down-weighting of severe outliers can be theoretically justified since several of our outliers are in any case potentially suspect. For example, for smartphones, the maximum observed usage time is implausibly high at about 30 days, indicating continuous usage (with the display on) for the entire month\footnote{Comparatively, the 95\% percentile of smartphone usage time is only about 9.5 days.}. Clearly this measurement does not likely represent the actual usage time of a single user but instead might be a measurement error.

\begin{figure}[!htb]
\centering
\includegraphics[width=6.25in]{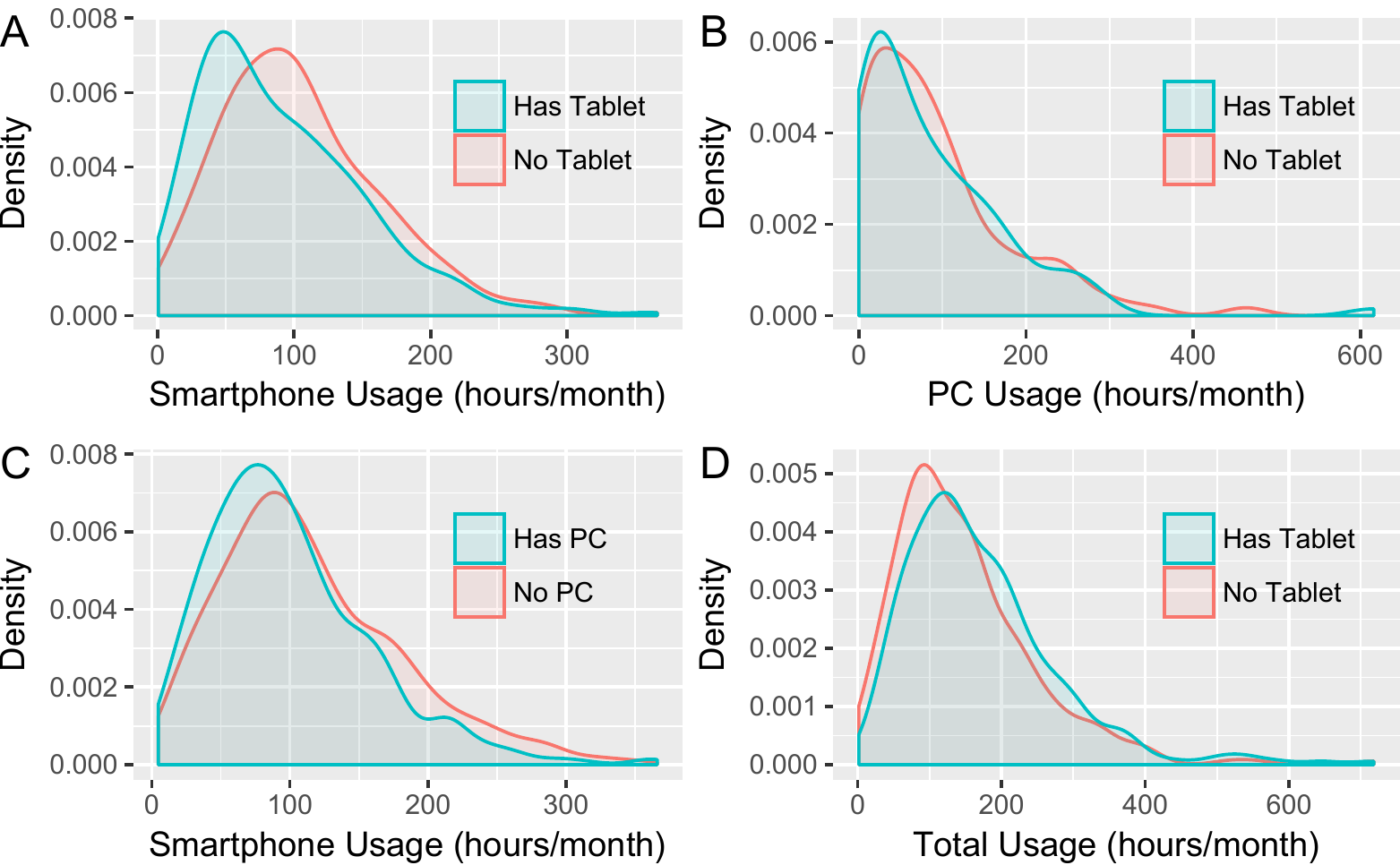}
\caption{A) Density estimates of smartphone usage time by tablet ownership (H1), B) Density estimates of PC usage time by tablet ownership (H2), C) Density estimates of smartphone usage time by PC ownership (H3), D) Density estimates of total device usage time by tablet ownership (H4)}
\label{h1h4_density}
\end{figure}

\begin{figure}[!htb]
\centering
\includegraphics[width=3.125in]{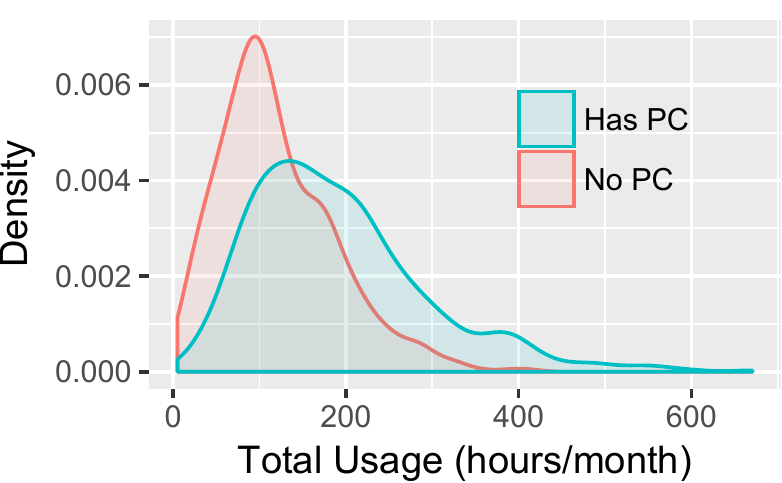}
\caption{Density estimates of total device usage time by PC ownership (H5)}
\label{h5_density}
\end{figure}

Finally, we perform the robust linear regression analyses. Table \ref{testing_results} details the results of each hypothesis testing including the estimated regression coefficient, statistical significance and whether these results support the hypothesis. Additionally, for illustration, Figures \ref{h1h4_density} and \ref{h5_density} detail the differences in distributions (kernel density estimates) that characterize each hypothesis test result.

For H1, we find that tablet ownership decreases smartphone usage by about 12.50 hours a month thus supporting H1 and the statement that tablets are a partial substitute for smartphones.

For H2, we find that tablet ownership does not significantly decrease PC usage thus refuting H2 and the statement that tablets are a partial substitute for PCs. Interestingly, this refutes the results in \citet[Section 5.2]{finley2016} which did not use a matching method. Furthermore, as we will show in the PC subtype testing in Section \ref{desktop_laptop_robustness}, this finding holds even when considering only laptop users (rather all PC users). Though overall, the sample size of the M2 matching (n=148) is still relatively small and, as we will discuss in Section \ref{device_sharing_robustness}, we are unable to perform device sharing testing for H2. Therefore, interpretations based on H2 results should still be performed with caution.

In terms of H3, we find surprisingly that PC ownership decreases smartphone usage by about 13 hours a month thus refuting H3 and the statement that PCs are not a partial substitute for smartphones. In other words, we find support for the statement that PCs are also a partial substitute for smartphones.

For H4 and H5, we find that both tablet and PC ownership significantly increases total device usage by about 20 and 57 hours a month respectively, therefore supporting H4 and H5 and the statements that tablet and PC ownership prompt additional (non-substituted) usage. The support for H4 backs up the findings of \citet{muller2015} based on self-reported electronic diaries. Though, as we will discuss in Section \ref{device_sharing_robustness}, device sharing testing for H4 suggests that the additional (non-substituted) tablet usage could be the result of device sharing within households. Finally, the support for H5 reinforces the findings of \citet{matthews2009} based on semi-structured interviews. The device sharing testing for H5 suggests that the additional (non-substituted) PC usage is not likely the result of device sharing within households. Additionally, the support for H5 also holds when considering laptop users and desktop users separately, though with desktop ownership increasing total device usage by 96 hours/month compared to 54 hours/month for laptop ownership.

Also in terms of diversity, the coefficient standard errors for H1, H3, and H4 are relatively large indicating substantial diversity in substituted and additional (non-substituted) usage between users. In other words, even though the coefficients are reasonable and descriptive point estimates, user diversity is also a part of the story. Similarly, user usage diversity has been observed extensively in previous mobile usage studies \citep{falaki2010,bohmer2011,soikkeli2013,finley2017a,hintze2017}.

\begin{table}[!htb]
\centering
\begin{threeparttable}
\caption{Hypotheses Testing Results}
\label{testing_results}
\begin{tabular}{lllll}
\toprule
{\bf Hypothesis} & {\bf Matching} & {\bf Depen. Variable} & {\bf Coefficients (hours/month)\tnote{a}} & {\bf Supported? }\\ \midrule
H1 & M1 & Smartphone Usage & -12.51 (4.99)$\ast{\ast}$ & Yes \\
H2 & M2 & PC Usage & -0.92 (11.94) & No \\ 
H3 & M3 & Smartphone Usage & -13.02 (3.26)$\ast{\ast}\ast$  & No \\
H4 & M4 & Total Device Usage & 19.58 (6.62)$\ast{\ast}$ & Yes \\
H5 & M5 & Total Device Usage & 57.43 (4.62)$\ast{\ast}\ast$ & Yes \\
\bottomrule
\end{tabular}
\begin{tablenotes}
    \scriptsize
    \item[a] Regression coefficients include standard errors and significance levels ($\ast$ : 5\%, $\ast\ast$ : 1\%, $\ast{\ast}\ast$ : 0.1\%).
\end{tablenotes}
\end{threeparttable}
\end{table}

\subsubsection{Device Sharing Testing}\label{device_sharing_robustness}
As mentioned, device sharing between members of a household is a potential concern in the testing of H2, H4, and H5. Therefore we also test these hypotheses with only one person household users (thus lessening the possibility of shared devices). Tables \ref{cem_results_device_sharing} and \ref{testing_results_device_sharing} in the appendix detail the matching and results of this testing respectively.

We find no significant difference for H5. However, H4 does show a significant difference (specifically the tablet ownership variable is no longer significant). Therefore for H4 we cannot exclude the possibility that the observed additional (non-substituted) tablet usage is the result of device sharing. Though, we note that the matching sizes for these one person household tests are (naturally) quite a bit smaller. Finally, we are unable to test H2 with one person household users due to too small a sample size\footnote{We note that we can apply the correlation approach (from \citet{finley2016} and Section \ref{correlation_approach}) to the subset. In this case the testing does not find statistically different results than the correlation approach with all households.} (24 users).

\subsubsection{PC Subtype (Desktop and Laptop) Testing}\label{desktop_laptop_robustness}
Additionally, we test the PC related hypotheses with the PC subtypes of desktop and laptop separately. Specifically, we test the hypotheses H2, H3, H5, with laptop-only users (69.15\% of PC users) and H3, H5 with desktop-only users (20.29\% of PC users). Unfortunately, we are unable to test H2 with desktop-only users due to too small a sample size (38 users). We exclude users with multiple PCs from the tests (7.6\% of PC users). Tables \ref{cem_results_pc_subtypes} and \ref{testing_results_pc_subtypes} in the appendix detail the matching and results of this testing respectively. We find no significant difference with the exception that, as previously discussed, desktop ownership appears to prompt more additional (non-substituted) usage than laptop ownership.

\subsubsection{Correlation Approach for H1-H3}\label{correlation_approach}
For comparison purposes, we also calculate the correlation coefficients between device type usage times for different device type ownership groups (for example, the correlation between tablet and PC usage times for the group of users with both a tablet and PC). This approach is the same as used in \citet{finley2016} (though \citet{finley2016} used a smaller dataset). While the regression approach answers the question {\it how does device type ownership affect the usage of another device type?}, the correlation approach answers the related question {\it how does device type usage affect the usage of another device type?}.

The correlation coefficients are calculated based on a robust correlation coefficient known as the OP correlation \citep{wilcox2017}. The OP correlation skips extreme outliers in the correlation calculation. The estimate of the OP correlation coefficient and significance level is performed through a percentile bootstrap method (with 1000 bootstraps) that is robust to heteroscedasticity \citep{wilcox2017}. Table \ref{correlation_results} details these correlations and their significance levels. 

As expected the correlations support the conclusions (H1-H3) of the regression analysis. In comparison to \citet{finley2016}, we find a correlation between Tablet and PC usage of 0.01 compared to -0.24 in \citep{finley2016}. In other words, we still find a significant difference between \citet{finley2016} and the current study even when using the same approach (without matching). The exact reason for this difference is difficult to pinpoint, though the sample size in \citet{finley2016} was only 52 users compared to 377 in this analysis.

\begin{table}[!htb]
\centering
\begin{threeparttable}
\caption{Correlation Approach Testing Results for H1-H3}
\label{correlation_results}
\begin{tabular}{llllll}
\toprule
{\bf Hypothesis} & {\bf User Subset\tnote{a}} & {\bf Users} & {\bf Corr Variables\tnote{a}} & {\bf Corr\tnote{b}}  & {\bf Supported? }\\ \midrule
H1 & Has S and T & 843 & S and T Usage Times & -0.12$\ast{\ast}\ast$ [-0.20 -0.04]& Yes\\
H2 & Has T and PC & 377 & T and PC Usage Times & 0.01 [-0.11  0.15] & No\\
H3 & Has S and PC & 2610 & S and PC Usage Times & -0.12$\ast{\ast}\ast$ [-0.16, -0.08] & No\\ 
\bottomrule
\end{tabular}
\begin{tablenotes}
    \scriptsize
    \item[a] S=Smartphone, T=Tablet.
    \item[b] Correlation coefficients include significance level ($\ast$ : 5\%, $\ast\ast$ : 1\%, $\ast{\ast}\ast$ : 0.1\%) and 95\% confidence interval estimated through a percentile bootstrap method \citep{wilcox2017} (1000 bootstraps).
\end{tablenotes}
\end{threeparttable}
\end{table}

\section{Discussion}\label{discussion}
Overall, we find support for device type substitution between smartphone and both tablets and PCs. 

Regarding smartphones and tablets, several theoretical factors support such substitution. For example, tablets generally run the same or only slightly modified smartphone applications, thus making substitution easier as the cost of learning a substitute program or alternative interface is small. Additionally, both smartphones and tablets are used extensively during the evening hours typically spent at home, thus providing the usage overlap needed for such substitution. However other factors inhibit substitution of certain usage. For example, smartphones smaller and more portable form factor compared to tablets naturally matches well to quick (often) informational glances known as micro-usage \citep{ferreira2014}. Therefore this type of usage is unlikely to be substituted to a tablet. Additionally, such micro-usage is more concentrated during the mid-day hours when tablets are not as available \citep{ferreira2014}.

In terms of smartphones and PCs, several of the factors supporting and inhibiting substitution between smartphone and tablets such as usage overlap and differences in form factor are similarly applicable. 

Finally, the lack of substitution between tablets and PCs is harder to explain theoretically. As we note in Section \ref{shared_devices}, tablets and PCs may be shared among several members of a household, therefore substitution effects might be more difficult to identify because usage is entangled between more users. Furthermore, we were unable to test H2 with only one person household users due to too small a sample size. Therefore, further research including qualitative research should be performed to elucidate the relationship between tablet and PC usage and provide more context. Additionally, the initial panel recruitment survey could be adjusted to include questions about the extent of device sharing within a household.

With regard to additional (non-substituted) usage, we find support for such usage with both tablets and PCs. Though, we cannot rule out that the additional tablet usage is the result of device sharing within households. In both the tablet and PC cases, theory suggests that such additional activities might be prompted by the ability to use the unique advantages of each device type. For example, the advantage of a large tablet display over a small smartphone display might prompt additional video sessions beyond the normal sessions that would occur with a smartphone. Towards this end, larger displays have been shown to increase user immersion in videos and games \citep{rigby2016,thompson2012} and user enjoyment in other tasks \citep{kim2011}.

\subsection{Limitations}\label{limitations}
There are several limitations of the study that should be noted.

As previously discussed, generalizability is a potential concern and limitation of our study. Though we study an overall diverse user group (both demographically and technographically), we still acknowledge differences between our active dataset and the US smartphone population in general (as detailed in Table \ref{panel_demos}). Additionally, as mentioned, the lack of support for monitoring Apple macOS means that Apple PC users are not included in the active dataset. Therefore generalizations to this specific user group are limited. These limitations should be considered when drawing general conclusions.

Relatedly, our analysis is limited to only a single month. Therefore inter-month variability in user usage may affect certain quantitative results. Though assuming such variability is asynchronized between users, such effects should remain small\footnote{Users are geographically dispersed across the US, thus negating local geographic synchronization.}. Furthermore, given a moderate panel churn rate, longitudinal user analyses face a trade-off between sample size and analysis period length\footnote{Though we note that Verto Analytics does use retention bonuses to try and help reduce churn.}. In this study, we use a relatively large sample size and short analysis period, whereas future studies could analyze different combinations (i.e. smaller sample size but longer analysis period).

Finally, as mentioned, display-off usage (such as listening to music) is not included in the analysis. Therefore the results are limited to display-on usage substitution. Additionally, device type substitution where display-off usage (of one device) is substituted for display-on usage (of another device) or vice versa could affect device substitution estimates. Specifically, additional (non-substituted) usage would be overestimated and the substituted usage underestimated or vice versa. Though we estimate this effect would be small due to the dominance of display-on usage, further research is needed to clarify fully. 

Additional limitations inherent to device-based monitoring studies, in general, are also applicable. For example, the analysis does not compensate for the time between when the user stops using the device and when the display turns off due to inactivity (assuming the user does not explicitly turn the display off). As discussed in \citet{hintze2017}, this is a limitation of all similar studies as typical devices cannot yet track user attention.

\subsection{Implications and Future Work}\label{implications}
Regarding result implications, we briefly discuss the implications for mobile advertisers and ad-driven application providers. 

As a background, for many mobile advertisers and application providers the atomic unit of mobile ad inventory is the impression\footnote{Impression inventory is often sold in units of 1000. An alternative inventory unit is the click where advertisers pay per click rather than per impression.}. An impression is the displaying of an ad\footnote{Often the ad is a banner at the top or bottom of the app.} within an application for a typically short length of time\footnote{The display time depends on the applications ad refresh rate.}. Therefore a primary determinate of an application provider's impression inventory is the total usage time for the application.

Thus, given the relationship between time usage and inventory, device type substitution patterns help in understanding inventory changes for different device types. For example, a further adoption of tablet devices (by smartphone users) might both increase tablet impression inventory at the expense of smartphone impression inventory and general additional tablet impression inventory. In other words, further adoption can both shift inventory and change the type of ad inventory (as smartphone and tablet ads are considered separate).

Beyond even these general understandings, the explicit quantification of such substitution can help in parameterizing future higher level research models such as a system dynamics model of the mobile content ecosystem. Such system dynamics models have been helpful in characterizing other parts of the mobile ecosystem such as digital service platform competition \citep{ruutu2017}, mobile voice diffusion and competition \citep{casey2012}, and mobile application services (stores) \citep{wang2016}. As \citet{casey2012} even admit, a common reason for not performing quantitative system dynamic modeling is a lack of data (used for initial parameterization and relationship specification). We look to explore such modeling in future work.

Regarding additional future work, an application level analysis of substitution and additional usage similar to the application level analysis in \citet{finley2017a} would be a natural next step. The primary challenge in such an analysis is reconciling the device type and platform differences in application names, categories, etc. especially in the case of smartphone and tablet versus PC. 

\section{Conclusions}\label{conclusions}
In this work, we provide estimates of device type substitution using device-based monitoring data and a robust paired matching method (to isolate the substitution effect). More specifically, the estimates allow the testing of five device type substitution hypotheses that span three different device types (smartphone, tablet, and PC). The results suggest that tablets and PCs are both partial substitutes for smartphones and yet also prompt significant additional (non-substituted) usage. Quantitatively, the estimates indicate that tablet ownership and PC ownership decrease smartphone usage by 12.5 and 13 hours/month, while prompting 20 and 57 hours/month of additional (non-substituted) usage respectively. Though we also find significant inter-user diversity in terms of these estimates. Finally, the results suggest that tablets are not a substitute for PCs despite the similarities between PCs (primarily laptops) and tablets in terms of portability and display size. Though this result is less robust and requires further study. Overall the results have implications, for example, for current mobile ecosystem players such as mobile advertisers and content providers and for parameterizing future research models.

\begin{acks}
The authors would like to thank Timo Smura, Heikki H{\"a}mm{\"a}inen, and Kalevi Kilkki for providing feedback on the manuscript.
\end{acks}

\bibliographystyle{ACM-Reference-Format}
\bibliography{main}
\vspace{3mm}
{\small Received November 2017; revised January 2018; accepted January 2018}

\appendix
\section{Appendix}
\label{App:AppendixA}

\begin{table}[!ht]
\centering
\begin{threeparttable}
\caption{M1 Pre-Match Difference Measures}
\label{prematching_comparison}
\begin{tabular}{lllllllll}
\toprule
{\bf Covariate } & {\bf Statistic\tnote{a} } & {\bf Type }  & {\bf L1 }  & {\bf Min }& {\bf 25\% }& {\bf 50\% }& {\bf 75\% }& {\bf Max }\\ \midrule 
  Has PC & 11.45 & (Cat) & 0.07 & - & - & - & - & - \\ 
  PC Usage Time & 10.56 & (Con) & 0.00 & 0.00 & 0.00 & 1.77 & 28.95 & 341.72 \\ 
  Display Size & 0.01 & (Con) & 0.00 & 0.00 & 0.00 & -0.10 & -0.30 & 0.30 \\ 
  Display Density & -12.64 & (Con) & 0.00 & 0.00 & -35.72 & -27.16 & 0.00 & 225.14 \\ 
  RAM & -0.07 & (Con) & 0.02 & -0.11 & 0.00 & 0.00 & 0.00 & 2.00 \\ 
  Display-to-Body Ratio & 0.00 & (Con) & 0.00 & 0.00 & 0.00 & 0.00 & 0.00 & 0.05 \\ 
  General Platform & 14.07 & (Cat) & 0.04 & - & - & - & - & - \\ 
  Age & -4.11 & (Con) & 0.05 & 0.00 & -4.00 & -4.00 & -6.00 & -2.00 \\ 
  Gender & 5.63 & (Cat) & 0.04 & - & - & - & - & - \\ 
  Marital Status & 8.42 & (Cat) & 0.05 & - & - & - & - & - \\ 
  Employment Status & 32.73 & (Cat) & 0.08 & - & - & - & - & - \\ 
  HH Income & 25.02 & (Cat) & 0.09 & - & - & - & - & - \\ 
  HH Size & 0.08 & (Con) & 0.03 & 0.00 & 0.00 & 0.00 & 0.00 & 2.00 \\ 
  Children in HH & 0.05 & (Con) & 0.01 & 0.00 & 0.00 & 0.00 & 1.00 & 3.00 \\ 
  Race & 3.23 & (Cat) & 0.02 & - & - & - & - & - \\ 
  Ethnicity & 2.30 & (Cat) & 0.01 & - & - & - & - & - \\ 
\bottomrule
\end{tabular}
\begin{tablenotes}
    \scriptsize
    \item[a] For continuous (Con) covariates the statistic is the difference in means, whereas for categorical (Cat) type covariates the statistic is the Chi\textsuperscript{2} test value.
\end{tablenotes}
\end{threeparttable}
\end{table}

\begin{table}[!ht]
\centering
\begin{threeparttable}
\caption{M1 Post-Match Difference Measures}
\label{postmatching_comparison}
\begin{tabular}{lllllllll}
\toprule
{\bf Covariate } & {\bf Statistic\tnote{a} } & {\bf Type }  & {\bf L1 }  & {\bf Min }& {\bf 25\% }& {\bf 50\% }& {\bf 75\% }& {\bf Max }\\ \midrule
  Has PC & 0.00 & (Cat) & 0.00 & - & - & - & - & - \\ 
  PC Usage Time & 0.88 & (Con) & 0.00 & 0.00 & 0.00 & 0.00 & 6.96 & -36.00 \\ 
  Display Size & -0.02 & (Con) & 0.02 & 0.00 & 0.00 & 0.00 & 0.00 & 0.04 \\ 
  Display Density & 1.91 & (Con) & 0.00 & 0.00 & 0.00 & 30.94 & 0.00 & 0.00 \\ 
  RAM & 0.00 & (Con) & 0.01 & 0.00 & 0.00 & 0.00 & 0.00 & 0.00 \\ 
  Display-to-Body Ratio & 0.00 & (Con) & 0.00 & 0.00 & 0.00 & 0.00 & 0.00 & 0.00 \\ 
  General Platform & 0.00 & (Cat) & 0.00 & - & - & - & - & - \\ 
  Age & -0.72 & (Con) & 0.00 & 0.00 & 1.00 & -2.00 & -1.00 & 1.00 \\ 
  Gender & 0.00 & (Cat) & 0.00 & - & - & - & - & - \\ 
  Marital Status & 2.41 & (Cat) & 0.02 & - & - & - & - & - \\
  Employment Status & 6.03 & (Cat) & 0.07 & - & - & - & - & - \\ 
  HH Income & 4.05 & (Cat) & 0.07 & - & - & - & - & - \\ 
  HH Size & 0.04 & (Con) & 0.03 & 0.00 & 0.00 & 0.00 & 0.00 & 1.00 \\ 
  Children in HH & 0.08 & (Con) & 0.04 & 0.00 & 0.00 & 0.00 & 0.00 & 0.00 \\ 
  Race & 1.12 & (Cat) & 0.01 & - & - & - & - & - \\ 
  Ethnicity & 3.62 & (Cat) & 0.02 & - & - & - & - & - \\ 
\bottomrule
\end{tabular}
\begin{tablenotes}
    \scriptsize
    \item[a] For continuous (Con) covariates the statistic is the difference in means, whereas for categorical (Cat) type covariates the statistic is the Chi\textsuperscript{2} test value.
\end{tablenotes}
\end{threeparttable}
\end{table}

\begin{table}[!ht]
\centering
\caption{CEM Matchings for Device Sharing Testing}
\label{cem_results_device_sharing}
\begin{tabular}{llll}
\toprule
{\bf Matching } & {\bf User Subset} & {\bf Treatment} & {\bf Users} \\ \midrule
M6 & One Person HH and Has PC & Has Tablet & 24 \\
M7 & One Person HH & Has Tablet & 84 \\
M8 & One Person HH & Has PC & 214 \\ 
\bottomrule
\end{tabular}
\end{table}

\begin{table}[!ht]
\centering
\begin{threeparttable}
\caption{Modeling Results for Device Sharing Testing}
\label{testing_results_device_sharing}
\begin{tabular}{lllll}
\toprule
{\bf Hypothesis} & {\bf Matching} & {\bf Depen. Variable} & {\bf Coef. (hours/month)\tnote{a}} & {\bf Supported? }\\ \midrule
H2 & M6 & PC Usage & -\tnote{b} & - \\
H4 & M7 & Total Device Usage & 18.75 (19.96) & No \\
H5 & M8 & Total Device Usage & 52.21 (10.39)$\ast{\ast}\ast$ & Yes \\
\bottomrule
\end{tabular}
\begin{tablenotes}
    \scriptsize
    \item[a] Regression coefficients include standard errors and significance levels ($\ast$ : 5\%, $\ast\ast$ : 1\%, $\ast{\ast}\ast$ : 0.1\%).
    \item[b] The matching size is too small for analysis.
\end{tablenotes}
\end{threeparttable}
\end{table}

\begin{table}[!ht]
\centering
\caption{CEM Matchings for PC Subtype (Desktop and Laptop) Testing}
\label{cem_results_pc_subtypes}
\begin{tabular}{llll}
\toprule
{\bf Matching } & {\bf User Subset} & {\bf Treatment} & {\bf Users} \\ \midrule
M9 & Has PC (Laptop) & Has Tablet & 104\\
M10 & All Users & Has PC (Laptop) & 996\\
M11 & All Users & Has PC (Laptop) & 1006\\
M12 & Has PC (Desktop) & Has Tablet & 38\\
M13 & All Users & Has PC (Desktop) & 390\\
M14 & All Users & Has PC (Desktop) & 396\\
\bottomrule
\end{tabular}
\end{table}

\begin{table}[!ht]
\centering
\begin{threeparttable}
\caption{Modeling Results for PC Subtype (Desktop and Laptop) Testing}
\label{testing_results_pc_subtypes}
\begin{tabular}{lllll}
\toprule
{\bf Hypothesis} & {\bf Matching} & {\bf Depen. Variable} & {\bf Coef. (hours/month)\tnote{a}} & {\bf Supported? }\\ \midrule
H2 & M9 & PC Usage (Laptop) & 3.95 (18.18) & No\\
H3 & M10 & Smartphone Usage & -16.30 (3.71)$\ast{\ast}\ast$ & No\\
H5 & M11 & Total Device Usage & 54.00 (5.03)$\ast{\ast}\ast$ & Yes\\
H2 & M12 & PC Usage (Desktop) & -\tnote{b} & - \\
H3 & M13 & Smartphone Usage & -15.94 (5.43)$\ast\ast$ & No\\
H5 & M14 & Total Device Usage & 95.89 (9.58)$\ast{\ast}\ast$ & Yes\\
\bottomrule
\end{tabular}
\begin{tablenotes}
    \scriptsize
    \item[a] Regression coefficients include standard errors and significance levels ($\ast$ : 5\%, $\ast\ast$ : 1\%, $\ast{\ast}\ast$ : 0.1\%).
    \item[b] The matching size is too small for analysis.
\end{tablenotes}
\end{threeparttable}
\end{table}
\end{document}